\documentclass[twocolumn,showpacs]{revtex4}

\usepackage{graphicx}
\usepackage{amsmath}
\usepackage{epsfig}
\usepackage{amssymb}
\usepackage{theorem}

\usepackage{hhline}

\newcommand{\ket}[1]{\left\vert#1\right\rangle}
\newcommand{\bra}[1]{\left\langle#1\right\vert}

\newcommand{\dmat}[2]{\ket{#1}\!\!\bra{#2}}
\newcommand{\beq}{\begin{equation}}
\newcommand{\eeq}{\end{equation}}
\newcommand{\bea}{\begin{eqnarray}}
\newcommand{\eea}{\end{eqnarray}}

\newcommand{\half}{\frac{1}{2}}

\newcommand{\tr}{\mbox{Tr}}

\newcommand{\3}{{\bf 3}}
\newcommand{\tb}{{\bf{\bar{3}}}}

\makeatletter
\def\btt#1{\texttt{\@backslashchar#1}}
\DeclareRobustCommand\bblash{\btt{\@backslashchar}}
\makeatother
\def\Bid{{\mathchoice {\rm {1\mskip-4.5mu l}} {\rm
{1\mskip-4.5mu l}} {\rm {1\mskip-3.8mu l}} {\rm {1\mskip-4.3mu l}}}}

\begin{document}

\title{Implications of Qudit Superselection rules for the Theory of
Decoherence-free Subsystems}
\author{Mark S. Byrd}
\email{mbyrd@physics.siu.edu}
\affiliation{Physics Department, Southern Illinois University, 
Carbondale, Illinois 62901-4401}


\begin{abstract}
The use of $d$-state systems, or qudits, in quantum 
information processing is discussed.  
Three-state and higher dimensional quantum systems are known to 
have very different properties from two-state systems, i.e., qubits.  
In particular there exist qudit states which 
are not equivalent under local unitary transformations 
unless a selection rule is violated.  This observation 
is shown to be an important factor in 
the theory of decoherence-free, or noiseless, subsystems.  
Experimentally observable 
consequences and methods for distinguishing these states 
are also provided, including the explicit construction of 
new decoherence-free or noiseless subsystems from qutrits.  
Implications for simulating quantum systems with quantum systems 
are also discussed. 
\end{abstract}


\maketitle




\section{Introduction}

At present it is not known which experiment will 
lead to the first reliable, prototypical quantum computing device.  
Quantum systems with two states, called qubits, 
are taken to be the basic unit for quantum information 
processing and storage.  However, in practice, these 
two states are often only two of a larger set of states.  
Therefore, one may wonder if a higher-dimensional system will 
eventually be used in its entirety for quantum computing.  
Higher dimensional quantum systems, 
which contain $d$ orthogonal states (called $d$-state systems 
hereafter), have many interesting 
properties which differ from those systems which have 
$d=2$ and may have advantages for quantum information processing.  
For example two three-state systems, or qutrits, can be more 
entangled than two qubits \cite{Caves/Milburn:99,Rungta/etal:00,qutritent}.  
$d$-state systems can also share a larger fraction of 
their entanglement \cite{Wootters:dqudits}.  

In addition to the differences in entanglement properties for 
quantum systems with more than two orthogonal 
states, there are differences 
in the selection rules governing the transitions between states.  
Some of these selection rules 
are referred to as ``superselection" rules.  
(See \cite{Kitaev/etal:ss} and references therein.)   
In the present article, 
a superselection rule will be taken to mean that a system's 
``principal" quantum numbers cannot be changed in a 
closed system.  A principal quantum number is defined 
here as one that identifies an irreducible 
representation (irrep) of a group.  
An example of such a rule is the preservation of the 
principal quantum number $j$ which applies when $J^2$ 
is a constant of the motion.  The differences 
in selection rules 
arise, in part, from the fact that for systems 
with $d\geq 3$, more than one principle 
quantum number is required.

Selection rules, including superselection rules, 
play an important role in quantum theory 
\cite{Wick/etal:52,Aharanov/Susskind,Wick/etal:70}.  
They often define 
a set of physically accessible states within a particular 
experiment.  Superselection rules could have important consequences 
for some quantum information processing protocols 
\cite{Verst/Cirac:03,Bartlett/Wiseman:03,Mayers:02,Kitaev/etal:ss,Bartlett/etal:04}.  
However, in the case of quantum cryptographic 
protocols, it has been shown that superselection rules do 
not aid in their security since these rules can, 
in principle, be violated \cite{Kitaev/etal:ss}.  
Here, the importance of selection 
rules for quantum information processing in other realms 
is explored.  We will see that selection rules 
have important implications for the theory of 
decoherence-free, or noiseless, subsystems (DFS/NS) 
\cite{Zanardi:97c,Duan:98,Lidar:PRL98,Knill:99a,Kempe:00,Lidar:00a} 
(for recent reviews see \cite{Lidar/Whaley:03,Byrd/etal:pqe04}), 
a topic which was also mentioned in connection with 
superselection rules in \cite{Bartlett/etal:04}.  

A DFS/NS can be described by a set of 
selection rules which are obeyed by a system-bath interaction.  
To compute on a DFS/NS, one violates the system-bath selection 
rule by using externally applied controls.  
Taking advantage of these selection rules and encoding in a DFS/NS 
has been shown to enable universal quantum 
computing on a noiseless subspace 
using a limited set of interactions.  
For example for certain DFS/NSs, the Heisenberg exchange 
interaction alone can be universal 
\cite{Bacon:Sydney,DiVincenzo:00a,Levy:01,Kempe:00,Lidar/Wu:01,Byrd/Lidar:ss}.  
This is quite an advantage in those 
systems which have readily available exchange interactions, 
but no other gating operations which are able to be 
easily or quickly performed.  
DFS/NSs also show promise for error protection and universal 
computing when combined with other methods.  
(See \cite{Byrd/etal:pqe04} for a review.)  One way 
to do this is to 
use dynamical decoupling to eliminate the noise on DFS/NS 
encoded qubits \cite{Wu/etal:02,Byrd/etal:05}.  Such 
``leakage elimination operations'' \cite{Wu/etal:02,Byrd/etal:05} 
can be used to prevent 
coupling of the two-state system whether it is 
a subspace of an $d$-state system or a logical qubit comprised 
of a subspace of a set of physical qubits.

In the case of leakage elimination and also 
the elimination of gating errors in a multistate system 
\cite{Tian:00}, controls are used to eliminate 
interactions between a subsystem (usually a physical or encoded 
qubit) of a multilevel system and other states in the system.  
The objective in this 
article is to explore the possibility of using 
the entire $d$-state system for quantum information processing.  
Computing with physical $d$-state 
was discussed in Refs.~\cite{Bullock/etal:05,Brennen/etal:05}  
and distillation protocols for physical $d$-state systems were 
discussed in Ref.~\cite{Bombin:05}.  
Both of these articles deal with quantum 
computing using $d$-state systems while 
the present article concerns encoding quantum information into 
collective DFS/NS using $d$-state systems.  
This is done by first describing a connection between quantum 
selection rules and operator algebras with group representation 
theory, operator algebras being useful for the description 
of DFS/NSs \cite{Knill:99a}.  Then, as mentioned 
above, implications of this for DFS/NS theory 
are examined.  

More specifically, this article is organized as follows.  
Section \ref{sec:gpth} contains conventions and 
labeling which will provide a basis 
for the group-theoretical treatment 
in this article including the definition of different types 
of selection rules.  Section \ref{sec:dfss} 
reviews some formal aspects of decoherence-free, or 
noiseless, subsystems.  Conventions for the choice 
of basis and eigenvalues for three-state systems are 
provided in Section \ref{sec:3sts}.  
These results will then 
be used in Section \ref{sec:3stdfss} 
for the construction of DFS/NS 
from systems having more than two orthogonal 
states.  This includes details of the 
a decoherence-free, or noiseless, qubit 
which is constructed from three qutrits and is immune to 
arbitrary collective errors.  
The properties which are important for the 
generalization to higher-dimensional systems are also 
important for the simulation of quantum systems with 
quantum systems.  Simulations and future work are 
discussed in the concluding section, Section \ref{sec:concl}.  
Two appendices provide some group-theoretical 
definitions, properties of young tableau, singlet 
states and a basis for $3\times 3$ matrices 
which are used in the text.


\section{Background}

\label{sec:gpth}

In Appendix \ref{app:gpth}, several definitions are given 
which are required in much of the rest of the article.  
These definitions can also be found Cornwell 
\cite{Cornwell:84I+II} (with slight differences).  The 
comments in Appendix \ref{app:gpth} 
are added to provide some extra explanation and 
context.  For our purposes, it is enough to note that 
there exist two inequivalent, fundamental, irreducible 
representations of $SU(d)$ for all $d\geq 3$.  
Definitions of ``inequivalent representations'' and 
``fundamental representations'' are two of the 
definitions provided in Appendix \ref{app:gpth}.  

For $SU(3)$, the two different fundamental 
irreps for $SU(3)$ will be denoted by $\3$ 
and $\tb$.  In general, representations will be 
denoted by bold-faced numbers with the numbers 
indicating the dimension of the representation.  
States within these two 
irreps will be denoted $\ket{i}$ and $\ket{\bar{i}}$ respectively.  
Tensor products will be written, for example, as $\ket{ii}$ 
($=\ket{i}\otimes\ket{i}$), 
$\ket{i\bar{i}}$ ($=\ket{i}\otimes\ket{\bar{i}}$), etc.


\subsection{Labels for Irreps}

For physical systems, a complete set of labels for the states 
is quite important since a complete set of labels is required 
in order to distinguish elements of a complete 
set of mutually orthonormal 
states.  In this section, such labels are discussed 
generally and then given explicitly 
for irreducible representations of $SU(2)$.  In 
Sec.~\ref{sec:3sts}, further discussion of this point is 
taken up and explicit labels for $SU(3)$ are given.  

Let $U$ be an element of a matrix representation of a Lie group, 
parameterized by a set of parameters 
$a_i$; $U=U(a_i)$.  
The elements of the matrix may then be denoted:
\beq
D^{(r)}_{m,m^\prime}(a_i) = \bra{r,m}U(a_i)\ket{r,m^\prime}.  
\eeq
In general, for $d\geq 3$ $r$ will represent more than 
one number.  Similarly, $m$ and $m^\prime$ will each 
represent more than one number.  
Quantum numbers $r$ represent ``principal quantum numbers'' 
and the quantum numbers $m,m^\prime$ will be referred to as 
``secondary quantum numbers.''

Let us give the familiar example of angular momentum in 
quantum mechanics.  The principal quantum number is 
taken to be $j$ which labels the total angular 
momentum through its relationship with the eigenvalue of the 
total angular momentum operator, $J^2$:
\beq
J^2\ket{j,m} = j(j+1)\ket{j,m}.
\eeq
If Euler angles $\alpha,\beta,\gamma$ are chosen to parameterize 
the matrix $U$, then the matrix elements are given by 
$$
D^{(j)}_{m,m^\prime}(\alpha,\beta,\gamma) = 
         \bra{j,m}U(\alpha,\beta,\gamma)\ket{j,m^\prime}.
$$
Here $j$ is the principal quantum number.  (Here there is only one.) 
And $m,m^\prime$ label states within an irrep.  Transitions 
may occur which change the $z$ component of the angular momentum 
$m$, but if $J^2$ is a constant of the motion, then 
$j$ will not change.  

These labels let us define a superselection rule as being one 
for which the principal quantum numbers are conserved; i.e., 
a superselection rule exists-and is not violated-if 
one cannot transform a state in one irreducible representation 
to a state in different irreducible representation.  
If two such representations are accessible and 
equivalent, we can include another ``principal quantum 
number'' to label this degeneracy.


\subsection{Superselection rules}

To make connection 
with previous work, note that the definition of a superselection 
rule used in this article is not significantly 
different from the one used by \cite{Kitaev/etal:ss} and 
\cite{Bartlett/etal:04} which state that a local superselection 
rule exists if there is a symmetry in the system.  In other words, 
a superselection rule exists 
if the system has the property that it is invariant under a 
group of transformations, ${\cal G}$-viz.,
\beq
\dmat{\psi}{\psi} = \int_{\cal G} U \dmat{\psi}{\psi} U^\dagger dU,
\eeq
where $U\in {\cal G}$ and $dU$ is the group-invariant Haar measure.  
The existence of such a symmetry 
implies that the group of all transformations 
on the Hilbert space is reducible.  This divides the space into 
superselection sectors.  Here we identify each superselection 
sector with an irrep of a group.  By our definitions, a 
superselection rule prevents the system from being transformed 
from a state within one sector to a state within another.


\section{Decoherence-Free or Noiseless Subsystems}

\label{sec:dfss}

In this section, a brief review of DFS/NS 
is provided using the notation 
of Refs.~\cite{Kempe:00} and \cite{Byrd/etal:05}.  This is 
followed by a statement of a theorem which apparently 
has not been previously 
applied to DFS/NS theory and which formally relates 
group theoretical representation theory to algebraic 
representation theory.  These and the example in the 
next section provide an application of these 
methods to a DFS/NS which is known.  This section 
is then followed by new results.  


\subsection{Definitions DFS/NS}

Consider a system $S$ coupled to a bath $B$ via the Hamiltonian 
\beq
H = H_S\otimes \Bid_B + \Bid_S \otimes H_B + H_I,
\eeq
where $H_S$ acts only on the system Hilbert space ${\cal H}_S$, 
$H_B$ acts only on the bath Hilbert space ${\cal H}_B$, 
$\Bid_S$ is the identity operator on the system Hilbert space, 
$\Bid_B$ is the identity operator on the bath Hilbert space,
and $H_I$ is the 
interaction Hamiltonian which acts on both the system and 
bath Hilbert spaces ${\cal H}_S\otimes {\cal H}_B$ 
and couples the two.  In general, $H_I$ can be written 
as a sum of operators which act on the system 
($S_\alpha$) and operators which act on the bath ($B_\alpha$),
\beq
H_I = \sum_\alpha S_\alpha \otimes B_\alpha.
\eeq
If there is no interaction Hamiltonian-i.e., when 
$H_I=0$-the system and bath evolve separately and unitarily:
\beq
U(t) = \exp[-iH_St]\otimes \exp[-iH_Bt],
\eeq
where $\hbar =1$.  

Consider the 
(associative) algebra, denoted ${\cal A}$, and 
generated by $H_S$ and the set of $S_\alpha$.  
This is a $\dagger$-closed algebra 
(if $A_i\in {\cal A}$, then  $A_i^\dagger \in {\cal A}$) which 
is, by assumption, reducible.  This implies that the algebra is 
isomorphic to a direct sum of $d_J\times d_J$ complex matrix 
algebras, each with multiplicity $n_J$:
\beq
\label{eq:Adef}
{\cal A} \cong \underset{J\in {\cal J}}{\oplus}\Bid_{n_J}\otimes 
                                     {\cal M}(d_J,\mathbb{C}).
\eeq
${\cal J}$ is a finite set labeling the irreducible components of 
${\cal A}$, and ${\cal M}(d_J,\mathbb{C})$ denotes a $d_J\times d_J$ 
complex matrix algebra.  
The commutant ${\cal A}^\prime$ of ${\cal A}$ is defined by 
\beq
{\cal A}^\prime = \{X:[X,A]=0, \; \forall \; A \in {\cal A}\}.  
\eeq
The elements of this set also form a $\dagger$-closed algebra.  
This algebra is also reducible, with 
\beq
{\cal A}^\prime \cong \underset{J\in {\cal J}}{\oplus}
                         {\cal M}(n_J,\mathbb{C})\otimes\Bid_{d_J}.
\eeq
An element of ${\cal A}$ can be written in block-diagonal form 
with $J$ denoting subblocks given in Eq.~(\ref{eq:Adef}).  
A typical block with given $J$ can be further decomposed as 
\begin{equation}
\label{eq:dfsmatrix}
\left[
\begin{tabular}{ccccccccccc}
\cline{1-3}
\multicolumn{1}{|c}{} &  &  & \multicolumn{1}{|c}{} &  &  &  &  &  &  &  \\
\multicolumn{1}{|c}{} & $M_{\alpha }$ &  & \multicolumn{1}{|c}{} &  &  &  &
&  &  & $\lambda =0$ \\
\multicolumn{1}{|c}{} &  &  & \multicolumn{1}{|c}{} &  &  & $\mu $ &  &  &
&  \\ \cline{1-6}
&  &  & \multicolumn{1}{|c}{} &  &  & \multicolumn{1}{|c}{$0$} &  &  &  &
\\
&  &  & \multicolumn{1}{|c}{} & $M_{\alpha }$ &  & \multicolumn{1}{|c}{$
\vdots $} &  &  &  & $\lambda =1$ \\
&  &  & \multicolumn{1}{|c}{} &  &  & \multicolumn{1}{|c}{$d_{J}-1$} &  &  &
&  \\ \cline{4-6}
&  & $\mu ^{\prime }:$ & $0$ & $\cdots $ & $d_{J}-1$ & $\ddots $ &  &  &  &
\\ \cline{8-10}
&  &  &  &  &  &  & \multicolumn{1}{|c}{} &  &  & \multicolumn{1}{|c}{} \\
&  &  &  &  &  &  & \multicolumn{1}{|c}{} & $M_{\alpha }$ &  &
\multicolumn{1}{|c}{$\lambda =n_{J}-1$} \\
&  &  &  &  &  &  & \multicolumn{1}{|c}{} &  &  & \multicolumn{1}{|c}{} \\
\cline{8-10}
\end{tabular}
\right]
\end{equation}
Here $\lambda$ labels the different degenerate subblocks, 
$0\leq\lambda \leq n_J-1$ and $\mu$ labels the states inside each 
$\lambda$-subblock, $0\leq \mu\leq d_J-1$.  
Associated with this decomposition of the algebra ${\cal A}$ is the
decomposition of the system Hilbert space:
\begin{equation}
{\cal H}_{S}=\sum_{J\in {\cal J}}\mathbb{C}^{n_{J}}\otimes \mathbb{C}^{d_{J}}.
\label{eq:repspc}
\end{equation}

Decoherence-free or noiseless subsystems can now be defined.  Let 
$\{\ket{\lambda_\mu}\}$, denote a subspace of ${\cal H}_S$ with 
given $J$.  Then the condition for the existence of an irreducible 
decomposition is 
\beq
A_\alpha\ket{\lambda_\mu} = 
  \sum_{\mu^\prime =1}^{d_J}M_{\mu\mu^\prime,\alpha}\ket{\lambda_{\mu^\prime}}
\eeq
for all $A_\alpha$, $\lambda$, and $\mu$.  Notice that 
$M_{\mu\mu^\prime,\alpha}$ does not depend on $\lambda$.  
Thus for a fixed $\lambda$, the subspace spanned by $\ket{\lambda_\mu}$ 
is acted upon in some nontrivial way.  However, because 
$M_{\mu\mu^\prime,\alpha}$ is not dependent on $\lambda$, each 
subspace defined by a fixed $\mu$ and running over 
$\lambda$ is acted upon in an identical manner by the decoherence 
process.  The information is stored in blocks with the same $J$, 
but different $\lambda$, and this defines a DFS/NS.  
Therefore, the labels which define the decoherence-free, 
or noiseless, states are the $\lambda$.  

A decoherence-free or noiseless {\it subspace} is one for which 
the matrices $M_{\mu\mu^\prime,\alpha}$ are one by one, i.e., 
they are numbers.  If the $M$ are numbers ($1\times 1$ matrices), 
then they act on a $1\times 1$ representation, which is necessarily 
a singlet state (a one-dimensional representation).  


\subsection{Weyl's unitary trick}

In the following sections, group theoretical methods will be 
used to identify DFS/NSs.  In particular, the representation theory 
of $SU(d)$ will be used repeatedly in order to find degeneracies 
which are able to represent DFS/NSs.  It is not immediately obvious 
that there exists an equivalence between group representation 
theory and the algebraic representation theory above.  
In other words, it may not 
be clear that the representation theory of the algebra of 
complex matrices is directly related to the theory of representations 
of the unitary groups.  
However, the representations are directly related and 
the use of this relation is sometimes 
referred to as Weyl's unitary trick.  
The theorem from Huang \cite{Huang:repbook} is stated here 
without proof.  (For a proof, see \cite{Huang:repbook}.)  

{\it Weyl's unitary trick.}
The following sets of representations on finite vector spaces 
are in one to one correspondence.  Moreover, under the 
correspondence, invariant subspaces and equivalences are 
preserved.
\begin{itemize}
\item[(i)]  holomorphic representations of $SL(d,\mathbb{C})$; 
\item[(ii)]  representations of $SL(d,\mathbb{R})$
\item[(iii)]  representations of $SU(d,\mathbb{R})$
\item[(iv)]  representations of $sl(d,\mathbb{R})$
\item[(v)]  representations of $su(d)$
\item[(vi)]  complex linear representations of $sl(d,\mathbb{C})$
\end{itemize}
Here a holomorphic (analytic) 
representation of $SL(d,\mathbb{C})$ is 
defined to be a homomorphism which is also a holomorphic map.  
The lowercase letters designate an algebra rather than a group.  
For example, $su(2)$ is the Lie algebra of the group $SU(2)$.

As an application of this ``trick,'' let us consider collective 
decoherence effects.  These are noises which act identically 
on every physical state in the system.  Let $\pi(L^\beta)$ 
a representation of a basis element of an abstract algebraic 
element $L^\beta$.  Let $\ket{a_i}$ and their tensor products 
carry a representation of a group $G$ generated by 
the set $\{L^\beta\}$, denoted ${\cal L}$.  Then 
\begin{gather}
\pi_e(L^\beta)(\ket{a_1}\otimes\ket{a_2}\otimes \cdots \ket{a_m})
\phantom{mmmmmmmmmmmmmmmmmmmmmmmmmmmmmmmmmmmmm} \nonumber \\
  \phantom{i} = [\pi_1(L^\beta)\ket{a_1}]\otimes(\ket{a_2}\otimes\cdots\ket{a_m}) 
          \nonumber \\
\phantom{mmi}  + \ket{a_1}\otimes[\pi_2(L^\beta]\ket{a_2})\otimes\cdots \ket{a_m} 
	  \nonumber \\
   + \cdots \phantom{mmmmmmmmmm} \nonumber \\
   \phantom{mmmmmmii} + \ket{a_1}\otimes\ket{a_2}\otimes \cdots[\pi_m(L^\beta)\ket{a_m}],
\label{eq:genalgaction}
\end{gather}
where $\pi_e$ is the representation on the entire space and 
$\pi_i$ is the representation on the $i^{th}$ subsystem.  
The algebra acting this way corresponds to the quantum numbers 
being additive.  For collective decoherence on a set of 
$m$ physical qudits, each $\pi_i(L^\beta)$ is identical.  
This provides a correspondence between the algebraic elements 
$L^\beta$ acting on the group and the algebraic elements 
which act as noises on the states and establishes 
the relation between tensor products of representations and 
direct sums of representations.

To exemplify and clarify these statements, the 
three-qubit DFS/NS is reexamined in the following 
section.  This will show how to provide generalizations 
of these operators, and the corresponding DFS/NSs to 
$d$-state systems.


\subsection{Example: Three-qubit DFS/NS}

\label{sec:3qbdfs}

The review of the three-qubit noiseless 
subsystem will enable the introduction of some general techniques, 
including Young tableau, in a more familiar context. 
(Rules for using Young 
tableau are given in \cite{Biedenharn,Cornwell:84II} 
and briefly discussed in Appendix \ref{app:gpth}.)  
In this case a decoherence-free, or noiseless, 
subsystem is formed 
from two doublet states in the Hilbert space of 
three two-state systems.  This code 
protects a single two-state subspace, referred to as the 
encoded qubit, from collective errors.  

Let us use Young's tableau to find the doublets.  For qubits, 
$d=2$, so there are two possible numbers with which boxes of 
a Young diagram can be filled.  Let us consider the 
following example of one box: 
$$
\setlength{\arrayrulewidth}{.4pt}
\begin{tabular}{|c|}
\hline \phantom{ai} \\ \hline
\end{tabular}
$$
Filling this with either a 1 or a 2, implies
$$
\setlength{\arrayrulewidth}{.4pt}
\begin{tabular}{|c|}
\hline \,1\, \\ \hline
\end{tabular}\;\;\;
\setlength{\arrayrulewidth}{.4pt}
\begin{tabular}{|c|}
\hline \,2\, \\ \hline
\end{tabular}
$$
This is a doublet, or two-dimensional representation of $SU(2)$.  
Taking the product gives
$$
\setlength{\arrayrulewidth}{.4pt}
\begin{tabular}{|c|}
\hline \phantom{ai} \\ \hline
\end{tabular}
\otimes 
\setlength{\arrayrulewidth}{.4pt}
\begin{tabular}{|c|}
\hline \phantom{ai} \\ \hline
\end{tabular} = \setlength{\arrayrulewidth}{.4pt}
\begin{tabular}{|c|}
\hline \phantom{ai} \\ \hline
 \phantom{ai} \\ \hline 
\end{tabular}
\oplus
\setlength{\arrayrulewidth}{.4pt}
\begin{tabular}{|c|c|}
\hline \phantom{ai} & \phantom{ai} \\ \hline
\end{tabular}.
$$
These can be filled in the following ways according 
to the rules for using Young's tableau.  The first can 
only have 
$$
\setlength{\arrayrulewidth}{.4pt}
\begin{tabular}{|c|}
\hline \,1\, \\ \hline
 \,2\, \\ \hline 
\end{tabular}
$$
The second can have
$$
\setlength{\arrayrulewidth}{.4pt}
\begin{tabular}{|c|c|}
\hline \,1\, & \,1\, \\ \hline
\end{tabular},\;\;\;
\setlength{\arrayrulewidth}{.4pt}
\begin{tabular}{|c|c|}
\hline  \,1\,& \,2\, \\ \hline
\end{tabular}\;\;\;
\setlength{\arrayrulewidth}{.4pt}
\begin{tabular}{|c|c|}
\hline \,2\, & \,2\, \\ \hline
\end{tabular}.
$$
which gives a singlet and a triplet respectively.  
This can be summarized in the equation 
${\bf 2} \otimes {\bf 2} = {\bf 3}\oplus {\bf 1}.$  
Taking the tensor products of three doublets,
\bea
\setlength{\arrayrulewidth}{.4pt}
\begin{tabular}{|c|}
\hline \phantom{ai} \\ \hline
\end{tabular}
\otimes 
\setlength{\arrayrulewidth}{.4pt}
\begin{tabular}{|c|}
\hline \phantom{ai} \\ \hline
\end{tabular}
\otimes 
\setlength{\arrayrulewidth}{.4pt}
\begin{tabular}{|c|}
\hline \phantom{ai} \\ \hline
\end{tabular} \!&=&\! \left(
\begin{tabular}{|c|}
\hline \phantom{ai} \\ \hline
 \phantom{ai} \\ \hline 
\end{tabular}
\oplus
\begin{tabular}{|c|c|}
\hline \phantom{ai} & \phantom{ai} \\ \hline
\end{tabular}\right)\otimes 
\begin{tabular}{|c|}
\hline \phantom{ai} \\ \hline
\end{tabular}                     \nonumber \\
   \!&=&\!
\setlength{\arrayrulewidth}{.4pt}
\begin{tabular}{|c|c|}
\hline \phantom{ai} & \phantom{ai}\\ \hline
 \phantom{ai} \\ \hhline{|-|~|}  
\end{tabular} 
\oplus
\setlength{\arrayrulewidth}{.4pt}
\begin{tabular}{|c|c|}
\hline \phantom{ai} & \phantom{ai}\\ \hline
 \phantom{ai} \\ \hhline{|-|~|}  
\end{tabular} 
\oplus 
\begin{tabular}{|c|c|c|}
\hline \phantom{ai} & \phantom{ai} & \phantom{ai}\\ \hline
\end{tabular}. \;\;\;\;\;\;\;
\eea
Filling in the numbers implies that there are two doublets and 
a quadruplet state in the direct sum decomposition, 
giving a total of eight states.  (Note that 
the set of three vertical boxes which one might have 
drawn here is not present.  This is because there is 
no nonzero state with three antisymmetric indices.)  

Now, the following convention is used for the 
computational basis states $\ket{0}$ and $\ket{1}$:
$|0\rangle =|1/2,1/2\rangle ,|1\rangle =|1/2, -1/2\rangle $. 
These are the two states of a single spin-$1/2$ particle, 
or a representation of the $j=1/2$ representation of 
$SU(2)$.  This convention is opposite to that of Ref.~\cite{Kempe:00}, 
but follows the conventions of 
\cite{Byrd/etal:05}, both of which provide more detail 
than is given here.  
Three-qubit DFS/NS encoded qubit will now be represented in 
the following way,
\begin{equation}
\left( 
\begin{array}{c}
(\left\vert 010\right\rangle -\left\vert 100\right\rangle )/\sqrt{2} \\ 
(\left\vert 011\right\rangle -\left\vert 101\right\rangle )/\sqrt{2} \\ 
(2\left\vert 001\right\rangle -\left\vert 010\right\rangle -\left\vert
100\right\rangle )/\sqrt{6} \\ 
(-2\left\vert 110\right\rangle +\left\vert 011\right\rangle +\left\vert
101\right\rangle )/\sqrt{6} \\ 
\left\vert 000\right\rangle \\ 
(\left\vert 001\right\rangle +\left\vert 010\right\rangle +\left\vert
100\right\rangle )/\sqrt{3} \\ 
(\left\vert 011\right\rangle +\left\vert 101\right\rangle +\left\vert
110\right\rangle )/\sqrt{3} \\ 
\left\vert 111\right\rangle%
\end{array}%
\right) \overset{\mbox{\LARGE{\phantom{X}}}}{%
\begin{array}{c}
{\bigg\}}\left\vert 0_{L}\right\rangle \\ 
{\bigg\}}\left\vert 1_{L}\right\rangle \\ 
{\mbox{\LARGE{${\Bigg\}}$}}}\mathcal{C}^{\perp }%
\end{array}%
}  \label{eq:3DFS}
\end{equation}%
With this notation $\left\vert 0_{L}\right\rangle =\alpha
_{0}(\left\vert 010\right\rangle -\left\vert 100\right\rangle )/\sqrt{2}%
+\beta _{0}(\left\vert 011\right\rangle -\left\vert 101\right\rangle )/\sqrt{%
2}$ (arbitrary superposition), and likewise $\left\vert 1_{L}\right\rangle
=\alpha _{1}(2\left\vert 001\right\rangle -\left\vert 010\right\rangle
-\left\vert 100\right\rangle )/\sqrt{6}+\beta _{1}(-2\left\vert
110\right\rangle +\left\vert 011\right\rangle +\left\vert 101\right\rangle )/%
\sqrt{6}$.  These states belong to the two $J=1/2$ 
irreps of $SU(2)$. The coefficients are Wigner-Clebsch-Gordan
coefficients \cite{Bohm:qmbook} and the last four states comprise a $J=3/2$
representation of $SU(2)$. The two $J=1/2$ irreps can be distinguished by a
degeneracy label $\lambda =0,1$. Thus a basis state in the eight-dimensional
Hilbert space is fully identified by the three quantum numbers $|J,\lambda
,\mu \rangle $, where $\mu $ is the $z$-component of the total spin $J$. In
this notation we can write $\left\vert 0_{L}\right\rangle =\alpha
_{0}|1/2,0,1/2\rangle +\beta _{0}|1/2,0,-1/2\rangle $ and $\left\vert
1_{L}\right\rangle =\alpha _{1}|1/2,1,1/2\rangle +\beta
_{1}|1/2,1,-1/2\rangle $.

If this encoded qubit is affected by collective errors-i.e., 
errors that act the same on each physical qubit-then no 
information is lost to the environment.  
The collective errors are formed from linear 
combinations of the operators
$S^\alpha= \sum_i\sigma^\alpha_i$: 
\beq
S = \sum_\alpha a_\alpha S^\alpha,
\eeq
where $\alpha=x,$ $y,$ or $z$ and the $i$ denotes the 
physical qubit $1$, $2$, or $3$.  The states within blocks 
$\ket{0_L}$ and $\ket{1_L}$ couple in exactly the same way, 
but neither block couples with states outside of that block.  
Logical operations create superpositions of these blocks.  
This can be described in terms of group 
representation theory, using Weyl's unitary trick.  
A basis for the algebra which spans the space of collective 
errors can be chosen to be the Lie algebra of $SU(2)$, $su(2)$.  
If we consider the action of the algebra on the entire space of 
the three qubits and suppose that this is a representation 
of $su(2)$ as well, then the representation on the entire 
space of three qubits is affected by the same operation 
on each physical qubit.  This is the statement made 
generally in Eq.~(\ref{eq:genalgaction}).  

For the example of three qubits, the matrix, 
Eq.~(\ref{eq:dfsmatrix}), can be found using the DFS/NS 
transformation [Eq.~(18) of \cite{Byrd/etal:05}].  
In the DFS/NS basis the explicit form is  given by
\begin{widetext}
\bea
S_{\mbox{\scriptsize dfs}} &=& %
U_{\mbox{\scriptsize dfs}}^{\phantom{-1}}SU_{\mbox{\scriptsize dfs}}^{-1} 
                               \nonumber \\
               &=& \left(\begin{array}{cccccccc} 
                  a_3 & a_1-ia_2 & 0 & 0 & 0 & 0 & 0 & 0 \\
                  a_1+ia_2 & -a_3 & 0 & 0 & 0 & 0 & 0 & 0 \\
   		  0  &  0 & a_3 & a_1-ia_2 & 0 & 0 & 0 & 0 \\
		  0 & 0 &  a_1+ ia_2 & -a_3 & 0 & 0 & 0 & 0 \\
		  0 & 0 & 0 & 0 & 3a_3 & \sqrt{3}(a_1-ia_2) & 0 & 0 \\
		  0 & 0 & 0 & 0 & \sqrt{3}(a_1+ia_2) & a_3 & 2(a_1-ia_2) & 0\\
		  0 & 0 & 0 & 0 & 0 & 2(a_1+ia_2) & a_3 & \sqrt{3}(a_1-ia_2)\\
		  0 & 0 & 0 & 0 & 0 & 0 & \sqrt{3}(a_1+ia_2) & -3a_3 
		  \end{array}\right). \;\;\;\;\;\;\;\;\;\;\;
\eea
\end{widetext} 
Thus we see that the two states of the logical zero transform 
in exactly the same way as the two states of the logical one 
under collective operations.  

In the context of this example, let us also 
consider a collection of physical qubits.  When collective 
errors occur, the form of these errors is 
\beq
\pi_e(L^\beta) = \sum_i \pi_i(L^\beta),  
\eeq
where $L^\beta$ is an element of the algebra and 
the subscript identifies the error as acting on the 
$i^{th}$ physical qubit.  
In other words, the operator $L^\beta$ acts on the 
system of qubits by acting with $L_1^\beta$ on qubit 1, 
$L_2^\beta$ on qubit 2, etc.  Each $L_i^\beta$ is 
identical, but acts on a different two-state system.  
Therefore, 
the statement that collective errors occur, is 
the statement that 
the entire system of qubits transforms as a representation 
of $su(2)$.  By the unitary trick, there is a 
direct correspondence between the representation theory 
of the group $SU(2)$ 
and the representation theory of the algebra with complex 
coefficients.  In this case, the $SU(2)$ transformation of 
the direct product of three two-dimensional representations 
is expressed as the direct sum of two three-dimensional 
representations (and a four-dimensional representation).  

This perspective of collective errors will now be used in 
the following sections to describe DFS/NSs of higher dimensional 
Hilbert spaces.  Though the arguments here are primarily 
restricted to the concrete example of $SU(3)$, they 
can readily be extended to any $SU(d)$.


\section{Types of Qutrit States and Labeling}

\label{sec:3sts}

In this section explicit labels are provided for the states 
of qutrits.  As noted in Appendix \ref{app:gpth} and also 
Sec.~\ref{sec:gpth}, two different types of qutrit states exist.  
This is true independent of the basis chosen for the algebra.  
Usually different sets of bases are 
chosen which reflect the symmetry 
of the physical system.  For example, 
given a representation of $SU(3)$, there are 
different subgroup chains which provide different 
possibilities for sets of measurements, corresponding 
to a different choice of basis elements,   
\begin{gather}
\label{chain1}
SU(3) \supset SU(2) \supset U(1)  \\
\label{chain2}
SU(3) \supset SO(3) \supset SO(2).
\end{gather}
The first is used in particle physics to describe 
the three lightest flavors of quarks \cite{Gell-Mann:64}.  
Associated with this subgroup chain is the set of 
Gell-Mann matrices.  The second of these subgroup 
chains is used in nuclear physics models, such as that 
of Elliott \cite{Elliott:58}.  The appropriate set 
of measurements depends on the ``good'' quantum numbers 
of the system.  

To each of these subgroup chains there corresponds a 
complete set of commuting operators (CSCO).  These operators are 
simultaneous observables which can be used to distinguish 
the different states within an irrep.  In the remainder 
of this article, subgroup chain (\ref{chain1}) is 
considered almost exclusively although the arguments 
can be applied to chain (\ref{chain2}) as well.  
By using this example, we 
are able to discuss the importance of 
the existence of two inequivalent fundamental irreps 
of the $SU(d)$ groups, both in the theory of DFS/NS 
and also in simulating quantum systems with quantum systems.


\subsection*{Labels}

As stated above, each of the two types 
irreps will be associated with the first subgroup chain 
(\ref{chain1}).  These will be called qutrit and 
barred states.  Qutrit states will be associated with the ${\3}$ 
representation and barred states 
will be associated with ${\tb}$, which is the complex 
conjugate of the ${\3}$ rep.  Throughout the rest of 
the article, states in the ${\tb}$ rep will have a bar 
over them to distinguish them from states in the 
${\3}$ rep: for example, the states 
$\ket{0},\ket{1},\ket{2}\in {\3}$ and 
$\ket{\bar{0}},\ket{\bar{1}},\ket{\bar{2}}\in {\tb}$.  

In order to provide a complete set of labels 
which distinguish two orthonormal states, a CSCO must be 
measured \cite{Bohm:qmbook}.  
For irreps of $SU(3)$, the following set of 
labels completely describe states within 
an irrep.  Each is associated with 
an operator in the CSCO.  Let $p,q$ label the irrep 
and $t$ label the 
eigenvalue of an $SU(2)$ subgroup of $SU(3)$; 
$T^2\ket{\psi} = t(t+1)\ket{\psi}$ for $\ket{\psi}$ an 
eigenstate of the operator $T^2$.  (Lowercase 
letters will represent the eigenvalues of the operators 
which will be denoted with an uppercase.)  The symbol $t_3$ will 
denote the eigenvalue of $T_3$, and $y$ will denote the eigenvalue 
of the operator $Y$.  In terms of the Gell-Mann matrices, 
(see Appendix \ref{app:alg})
\begin{gather} 
Y = \frac{1}{\sqrt{3}} \lambda_8, \nonumber \\
T^2 = \frac{1}{4}(\lambda_1^2+\lambda_2^2 +\lambda_3^2), \label{eq:eigenops}\\
T_3 = \half\lambda_3.  \nonumber 
\end{gather}
The quantum numbers $p$ and $q$ can be determined by the 
highest-weight states (described later in this section) 
or by measurement of the Casimir operators. 
(See Appendix \ref{app:alg}.) The Casimir operators, 
plus the set of operators in Eq.(\ref{eq:eigenops}) provide 
a CSCO.  States within any irreps can be written as 
$\ket{p,q,t,t_3,y}$, where $p$ and $q$ are assumed to be 
fixed and determined by the irrep.  To make a connection 
with the more familiar case of a spin-$j$ particle, $p,q$ 
should be considered ``principal" quantum numbers which 
label an irrep.  
When $J^2$ is a constant of the motion, then $j$ is 
fixed and cannot change.  
The analog for three-state systems is the 
conservation of the two quantum 
numbers $p$ and $q$.  It will be assumed here, 
unless otherwise stated, that $p$ and $q$ are both conserved.  
However, whether 
or not these quantum numbers are conserved in a particular 
experiment depends on the physical system in question.  

The comparison 
with the unitary representation of the group is made by labeling 
the unitary matrices in the following way:
$$
\bra{p,q;t,t_3,y}U\ket{p,q;t^\prime,t_3^\prime,y^\prime}  
          = D^{(p,q)}_{t,t_3,y;t^\prime,t_3^\prime,y^\prime},
$$
so that the matrix elements of $U$ are 
given by the functions $D$ and the rows (columns) are labeled 
by the primed (unprimed) numbers.  

For $SU(3)$ there are six raising and lowering operators which 
take one state in an irrep to another state in the same irrep.  
They are often denoted $U_\pm,V_\pm,T_\pm$.

\begin{figure}
\includegraphics{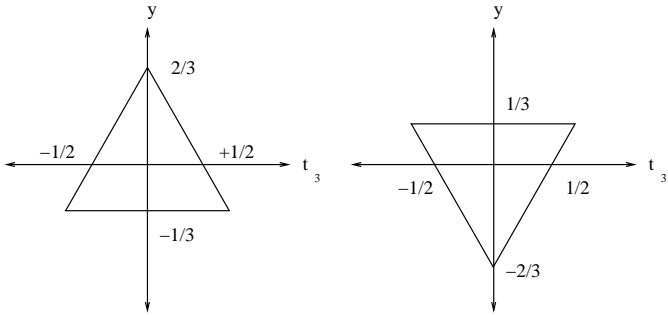}
\caption{\label{fig:fundirreps} State spaces for the 
${\bf 3}$ (left) and ${\tb}$ (right) reps.}
\end{figure}

Within an irrep, one can define a unique 
``maximum weight state'' (see \cite{Cornwell:84II,symmetry:book}).  
This state $\ket{\psi_m}$ is usually defined as the one for 
which the following relations hold:
\begin{equation}
\label{eq:maxstaterl}
T_+\ket{\psi_m}=0,\;\;\;V_+\ket{\psi_m}=0,\;\;\;U_-\ket{\psi_m}=0.
\end{equation}
Since the maximum weight state is unique for each irrep, it can 
be related to the labels, $p$ and $q$,  which identify the irrep,
\begin{equation}
\label{eq:pnq}
t_{3m} = \frac{p+q}{2},\;\;\;\; y_m = \frac{p-q}{3}.
\end{equation}
The examples of the two fundamental irreps ${\3}$ and ${\tb}$ 
are given in Fig.~\ref{fig:fundirreps} where the states 
are labeled according to the eigenvalues of $Y$ and $T_3$.  
The highest weight states in $\3$ and $\tb$ are the 
states with $t_3=1/2,y=-1/3$ and $t_3 =1/2,y=1/3$ 
respectively.

Let us now label the states $\ket{0},\ket{1},\ket{2}$ by 
using the full set of quantum numbers, 
\bea
\label{eq:3states}
\ket{0}&=& \ket{1,0,1/2,1/2,1/3}, \nonumber \\ 
\ket{1}&=& \ket{1,0,1/2,-1/2,1/3},\\ 
\ket{2}&=& \ket{1,0,0,0,-2/3}. \nonumber 
\eea
and 
\bea
\label{eq:3barstates}
\;\;\;\;\;\ket{\bar{0}}&=& \ket{0,1,-1/2,-1/2,-1/3},\nonumber \\ 
\ket{\bar{1}}&=& \ket{0,1,-1/2,1/2,-1/3},\\ 
\ket{\bar{2}}&=& \ket{0,1,0,0,2/3}.  \nonumber 
\eea

For future reference, note that an $SU(3)$ singlet state 
has the unique property that 
\begin{equation}
\label{eq:singletstaterl}
T_\pm\ket{\psi_s}=0,\;\;\;V_\pm\ket{\psi_s}=0,\;\;\;U_\pm\ket{\psi_s}=0.
\end{equation}
This follows from the fact that a ``singlet'' state is one for which 
there exists only one state within the irrep (see also 
Appendix~\ref{app:singlets} and the comment at the end of 
Sec.~III~A).


\section{Decoherence-free subspaces for three-state systems}

\label{sec:3stdfss}

In principle, one can find DFS/NSs from the formal theory 
provided in \cite{Knill:99a}.  However, until now, there has been 
no emphasis on the implications of the various irreps of a 
given group (with the exception of \cite{Bartlett/etal:04}).  
Here, in particular, it is shown that the 
distinction between the ${\bf 3}$ and ${\bar {\bf 3}}$ 
representations is tantamount to the 
identification and use of a DFS/NS for quantum error 
avoidance in qutrit systems.  

Again, in what follows three-state systems are used 
as explicit examples. However, the constructions here are 
readily generalizable to $SU(d)$.  


\subsection{Product states}

It is common in the quantum information literature to find statements 
such as ``... the maximally entangled state of two qutrits, 
$\ket{\phi}=\frac{1}{\sqrt{3}}(\ket{00}+\ket{11}+\ket{22})$.''  
However, we have just seen that there are two different irreducible 
fundamental representations of $SU(3)$.  In that case, we should 
distinguish between this state, $\ket{\phi}$ and 
$\ket{\phi^\prime}=\frac{1}{\sqrt{3}}(\ket{0\bar{0}}
                     +\ket{1\bar{1}}+\ket{2\bar{2}})$.
This observation has an important consequence since 
there is a striking difference between these two states.  
{\it The state $\ket{\phi^\prime}$ is an $SU(3)$ singlet, 
but $\ket{\phi}$ is not}.  
Singlet states are used in the theory of DFS in order to protect 
against all forms of errors.  The state $\ket{\phi}$ is not 
decoherence free.  
(To see that $\ket{\phi^\prime}$ is a singlet state, consult 
Appendix \ref{app:singlets}.)  

Note that, in principle, a local unitary can be found which will 
transform the state $\ket{\phi}$ into the state 
$\ket{\phi^\prime}$.  This implies that 
the amount of entanglement in $\ket{\phi}$ is 
the same as the amount of entanglement in $\ket{\phi^\prime}$.  
Therefore, whereas 
it has been conjectured that better quantum error correcting 
codes are those with more entanglement present in their states 
\cite{Scott:04}, no such correspondence can be made for DFS/NS.  
In other words, {\it there is no direct correspondence between 
the amount of entanglement in a DFS/NS and the efficacy, or 
error avoidance properties, of the encoded DFS/NS states.}  
This is indicated from the example just presented as well as the 
corresponding Young diagram wich applies to any two $d$-state 
systems.  

The states $\ket{\phi}$ and $\ket{\phi^\prime}$, 
belong to two different 
irreps.  Checking all quantum numbers shows that each of the 
two states has all secondary quantum numbers, $t,t_3,$ and $y$ 
equal to zero; that is, $t=0,t_3=0,$ and $y=0$.  
However, the primary quantum numbers differ.   
$\ket{\phi}$ has $p=1,q=0$ whereas 
$\ket{\phi^\prime}$, $p=0,q=1$.  This difference is experimentally 
measurable in 
several different ways.  One way is to 
find the highest weight state of the irrep through the use of the 
appropriate raising and/or lowering operators.  This will identify 
$p$ and $q$ through Eqs.~(\ref{eq:pnq}).  
Also, the differentiation between these states 
has implications for quantum error 
avoidance properties of the states.  
This provides another important method 
for experimental distinction.


\subsection{Young Tableau and DFSs}

\label{sec:youngtrits}

The use of Young's tableau proves very convenient for 
exploring the possibility of constructing collective 
DFS/NSs from qutrits (or qudits).  
When two or more irreducible representations 
occur in a tensor product of a set of states of any dimensions, these 
identical irreps will transform in the same way under an 
$SU(3)$ action on the entire space of physical subsystems 
and are therefore candidates for a DFS/NS. (See Section \ref{sec:dfss}).  
Let us now examine some tensor products of qutrits to determine 
the possibility of constructing DFS/NSs which are immune to 
collective errors.

The two different Young tableau for the ${\bf 3}$ and 
${\bf \bar{3}}$ representations are represented by 
$$
\setlength{\arrayrulewidth}{.4pt}
\begin{tabular}{|c|}
\hline \phantom{ai} \\ \hline
\end{tabular}
$$
which is filled with numbers $1,2,3$
$$
\setlength{\arrayrulewidth}{.4pt}
\begin{tabular}{|c|}
\hline 1 \\ \hline
\end{tabular}, \;\;\;
\setlength{\arrayrulewidth}{.4pt}
\begin{tabular}{|c|}
\hline 2 \\ \hline
\end{tabular}, \;\;\;
\setlength{\arrayrulewidth}{.4pt}
\begin{tabular}{|c|}
\hline 3 \\ \hline
\end{tabular},
$$
and 
$$
\setlength{\arrayrulewidth}{.4pt}
\begin{tabular}{|c|}
\hline \phantom{ai} \\ \hline
 \phantom{ai} \\ \hline 
\end{tabular}
$$
which is filled with numbers $1,2,3$ but is antisymmetric in the 
interchange of two rows.  This gives the following possibilities:
$$
\setlength{\arrayrulewidth}{.4pt}
\begin{tabular}{|c|}
\hline 1 \\ \hline
 2 \\ \hline 
\end{tabular}, \;\;\;
\setlength{\arrayrulewidth}{.4pt}
\begin{tabular}{|c|}
\hline 1 \\ \hline
 3 \\ \hline 
\end{tabular}, \;\;\;
\setlength{\arrayrulewidth}{.4pt}
\begin{tabular}{|c|}
\hline 2 \\ \hline
 3 \\ \hline 
\end{tabular}.
$$
These are the two inequivalent fundamental irreps.  

The tensor product of two ${\bf 3}$ gives the following 
$$
\setlength{\arrayrulewidth}{.4pt}
\begin{tabular}{|c|}
\hline \phantom{ai} \\ \hline
\end{tabular}
\otimes 
\setlength{\arrayrulewidth}{.4pt}
\begin{tabular}{|c|}
\hline \phantom{ai} \\ \hline
\end{tabular} = \setlength{\arrayrulewidth}{.4pt}
\begin{tabular}{|c|}
\hline \phantom{ai} \\ \hline
 \phantom{ai} \\ \hline 
\end{tabular}
\oplus
\setlength{\arrayrulewidth}{.4pt}
\begin{tabular}{|c|c|}
\hline \phantom{ai} & \phantom{ai} \\ \hline
\end{tabular}.
$$
The first is the ${\bf \bar{3}}$ rep and the second is a 
six-dimensional representation which can be shown by filling 
in the boxes with all possible symmetric combinations 
$$
\setlength{\arrayrulewidth}{.4pt}
\begin{tabular}{|c|c|}
\hline 1 & 1 \\ \hline
\end{tabular},\;\;\;
\setlength{\arrayrulewidth}{.4pt}
\begin{tabular}{|c|c|}
\hline 1 & 2 \\ \hline
\end{tabular},\;\;\;
\setlength{\arrayrulewidth}{.4pt}
\begin{tabular}{|c|c|}
\hline 1 & 3 \\ \hline
\end{tabular},\;\;\;
\setlength{\arrayrulewidth}{.4pt}
\begin{tabular}{|c|c|}
\hline 2 & 2 \\ \hline
\end{tabular},\;\;\;
\setlength{\arrayrulewidth}{.4pt}
\begin{tabular}{|c|c|}
\hline 2 & 3 \\ \hline
\end{tabular},\;\;\;
\setlength{\arrayrulewidth}{.4pt}
\begin{tabular}{|c|c|}
\hline 3 & 3 \\ \hline
\end{tabular}.
$$
To be precise, this is the ${\bf 6}$ rep.  The 
result of this can be written in the following 
equation: 
${\bf 3} \otimes {\bf 3} = {\tb} \oplus {\bf 6}$.  

Now, note that the result of the product of 
${\bf \bar{3}}$ and ${\bf 3}$ is given by
$$
\setlength{\arrayrulewidth}{.4pt}
\begin{tabular}{|c|}
\hline \phantom{ai} \\ \hline
 \phantom{ai} \\ \hline 
\end{tabular} \otimes 
\setlength{\arrayrulewidth}{.4pt}
\begin{tabular}{|c|}
\hline \phantom{ai} \\ \hline
\end{tabular} = 
\setlength{\arrayrulewidth}{.4pt}
\begin{tabular}{|c|c|}
\hline \phantom{ai} & \phantom{ai}\\ \hline
 \phantom{ai} \\ \hhline{|-|~|}  
\end{tabular} 
\oplus
\setlength{\arrayrulewidth}{.4pt}
\begin{tabular}{|c|}
\hline \phantom{ai} \\ \hline
 \phantom{ai} \\ \hline 
 \phantom{ai} \\ \hline 
\end{tabular}.
$$
The first tableau corresponds to an octet of states,
$$
\setlength{\arrayrulewidth}{.4pt}
\begin{tabular}{|c|c|}
\hline 1 & 1\\ \hline
 2 \\ \hhline{|-|~|}  
\end{tabular},\;\;\;
\setlength{\arrayrulewidth}{.4pt}
\begin{tabular}{|c|c|}
\hline 1 & 1\\ \hline
 3 \\ \hhline{|-|~|}  
\end{tabular},\;\;\;
\setlength{\arrayrulewidth}{.4pt}
\begin{tabular}{|c|c|}
\hline 1 & 2\\ \hline
 2 \\ \hhline{|-|~|}  
\end{tabular},\;\;\; 
\setlength{\arrayrulewidth}{.4pt}
\begin{tabular}{|c|c|}
\hline 1 & 2\\ \hline
 3 \\ \hhline{|-|~|}  
\end{tabular},\;\;\; 
\setlength{\arrayrulewidth}{.4pt}
\begin{tabular}{|c|c|}
\hline 1 & 3\\ \hline
 2 \\ \hhline{|-|~|}  
\end{tabular},\;\;\; 
\setlength{\arrayrulewidth}{.4pt}
\begin{tabular}{|c|c|}
\hline 1 & 3\\ \hline
 3 \\ \hhline{|-|~|}  
\end{tabular},\;\;\; 
\setlength{\arrayrulewidth}{.4pt}
\begin{tabular}{|c|c|}
\hline 2 & 2\\ \hline
 3 \\ \hhline{|-|~|}  
\end{tabular},\;\;\; 
\setlength{\arrayrulewidth}{.4pt}
\begin{tabular}{|c|c|}
\hline 2 & 3\\ \hline
 3 \\ \hhline{|-|~|}  
\end{tabular}. 
$$
The second corresponds to a singlet, as it can only be filled in 
one way,
$$
\setlength{\arrayrulewidth}{.4pt}
\begin{tabular}{|c|}
\hline 1 \\ \hline
 2 \\ \hline 
 3 \\ \hline 
\end{tabular}.
$$
Therefore, 
${\bf \bar{3}}\otimes {\bf 3} = {\bf 8} \oplus {\bf 1}$ 
\cite{octetnote}.  
This shows that the product of two three-dimensional 
representations of the same type do not give rise to a 
singlet state, whereas products of two reps 
of different types do give rise to a singlet state.  
Singlet states are decoherence-free since they are annihilated by 
all $SU(3)$ operators \cite{Lidar:PRL98}.  

Let us consider constructing a decoherence-free, or 
noiseless, qubit from qutrits.  We have already seen 
this is not possible using two identical 
qutrits, or an unbarred 
and a barred rep.  One may naturally ask about 
three unbarred (or three barred) states.  From the 
tableau, it can be shown that 
${\bf 3}\otimes {\bf 3}\otimes {\bf 3} = 
{\bf 8}\oplus {\bf 8} \oplus {\bf 1} \oplus {\bf 10}$.
This indicates that three qutrit states have a set of 
two degenerate reps.  This implies that a DFS/NS can 
be constructed with the two degenerate states representing 
the logical zero and logical one states of a qubit which 
is immune to collective noise.  

Note, however, that the product of a barred and two unbarred 
reps will have the following decomposition:  
$\tb\otimes {\bf 3} \otimes {\bf 3} = {\bf 15}\oplus {\bf 3} 
\oplus {\bf 3} \oplus {\bf \bar{6}}$.  
This indicates that one may also construct a DFS/NS from this 
set of states which can represent a decoherence-free qubit.  
Certainly these two are quite different subsystems.  The 
first has two degenerate eight-state subsystems and the 
second has two degenerate three-state systems.

In order to find the fewest number of physical qutrits 
which can be encoded such that a logical qutrit is 
protected from collective errors, four qutrits are 
taken: 
${\bf 3}\otimes {\bf 3}\otimes {\bf 3}\otimes {\bf 3} 
= {\bf 3}\oplus {\bf 3} \oplus {\bf 3} \oplus {\bf \bar{6}}
\oplus {\bf \bar{6}}\oplus {\bf 15}\oplus {\bf 15}\oplus {\bf 15}
\oplus {\bf \bar{15}}$.  In this case, a decoherence-free qutrit 
could be represented by three three-state systems and this is 
the smallest number of qutrit states which can represent 
such a qutrit DFS/NS.  

The analysis can be used for any $d$-state 
systems.  For example, one may ask for the least number 
of physical $d$-state systems which can be used to 
encode a logical qubit which is decoherence free with 
respect to collective errors.  
The answer can be found by again using Young tableau.  
{\it The tensor product of three $d$-state systems can 
be used to encode logical qubit into a NS/DFS.}  
This can be seen in the tableau of any ${\bf d}$ 
representation
$$
\setlength{\arrayrulewidth}{.4pt}
\begin{tabular}{|c|}
\hline \phantom{ai} \\ \hline
\end{tabular}.
$$
Taking the tensor product of three such systems, produces the following 
tableau
\bea
\label{tableau:3dfs}
\setlength{\arrayrulewidth}{.4pt}
\begin{tabular}{|c|}
\hline \phantom{ai} \\ \hline
\end{tabular}
\otimes 
\setlength{\arrayrulewidth}{.4pt}
\begin{tabular}{|c|}
\hline \phantom{ai} \\ \hline
\end{tabular}
\otimes 
\setlength{\arrayrulewidth}{.4pt}
\begin{tabular}{|c|}
\hline \phantom{ai} \\ \hline
\end{tabular} \!&=&\!
\left(
\begin{tabular}{|c|}
\hline \phantom{ai} \\ \hline
 \phantom{ai} \\ \hline 
\end{tabular}
\oplus
\begin{tabular}{|c|c|}
\hline \phantom{ai} & \phantom{ai} \\ \hline
\end{tabular}\right)\otimes 
\begin{tabular}{|c|}
\hline \phantom{ai} \\ \hline
\end{tabular} \nonumber \\
\!&=&\!
\setlength{\arrayrulewidth}{.4pt}
\begin{tabular}{|c|c|}
\hline \phantom{ai} & \phantom{ai}\\ \hline
 \phantom{ai} \\ \hhline{|-|~|}  
\end{tabular} 
\oplus
\setlength{\arrayrulewidth}{.4pt}
\begin{tabular}{|c|c|}
\hline \phantom{ai} & \phantom{ai}\\ \hline
 \phantom{ai} \\ \hhline{|-|~|}  
\end{tabular} 
\oplus
\setlength{\arrayrulewidth}{.4pt}
\begin{tabular}{|c|}
\hline \phantom{ai} \\ \hline
 \phantom{ai} \\ \hline 
 \phantom{ai} \\ \hline 
\end{tabular} \oplus 
\begin{tabular}{|c|c|c|}
\hline \phantom{ai} & \phantom{ai} & \phantom{ai}\\ \hline
\end{tabular}.\;\;\;\;\;
\eea
Therefore, {\it this is the smallest number of qudits for which a 
collective DFS/NS, 
representing a qubit in terms of qudits, exists.}  

The difference between a tensor product of two fundamental 
irreps which are equivalent and two which are not is clearly 
very important for constructing DFS/NS from higher dimensional 
systems.  The fact that one of the two states transforms 
differently than the other implies that a superselection rule 
which preserves the type of qutrit (or qudit) 
must exist in the system-bath 
interaction.  On the other hand, if one wants to create 
a DFS/NS by the use of decoupling controls according to the 
methods presented in \cite{Zanardi:98b,Viola:00a,Wu/Lidar:cdfs,Byrd/Lidar:ss,Viola:01a,Byrd/Lidar:01,Byrd/Lidar:ebb,Byrd/Lidar:pqe02,Lidar/Wu:02,Wu/etal:02,Zanardi:99a}, 
then one must recognize 
this as a quantum control problem in which the decoupling 
controls must provide the appropriate symmetry 
for those systems which do not otherwise 
obey the required superselection rule.  In other words, to create 
a DFS/NS from two inequivalent fundamental irreps, one 
must ensure that the appropriate transformation properties 
are obeyed.  Representing 
decoherence-free systems with $d$-state systems therefore 
requires knowledge of the transformation properties induced 
by experimental controls and system-bath interactions.  


\subsection{Three-qutrit DFS/NS}

As discussed in the previous section, 
Sec.~\ref{sec:youngtrits}, 
a noiseless subsystem can be formed from two octets in the 
Hilbert space of three three-state systems.  This 
logical qubit will be   
protected against arbitrary collective errors 
[see Eq.~(\ref{tableau:3dfs})].  

Using the conventions established by Eqs.(\ref{eq:3states}), 
the logical states can be given 
explicit labels according to the principal quantum numbers 
$p,q$ and eigenvalues of the operators $T$, $Y$, and $T_3$.  
The first of two octets will have a degeneracy label 
$0$, which indicates that it forms the logical 
zero state $\ket{0_L}$, 
\bea
\psi^{8,0}_1&=&(\ket{200}-\ket{020})/\sqrt{2}, \nonumber  \\
\psi^{8,0}_2&=&(\ket{100}-\ket{010})/\sqrt{2},  \nonumber  \\
\psi^{8,0}_3&=&(\ket{011}-\ket{101})/\sqrt{2},  \nonumber  \\
\psi^{8,0}_4&=&(\ket{211}-\ket{121})/\sqrt{2},  \nonumber  \\
\psi^{8,0}_5&=&(\ket{212}-\ket{122})/\sqrt{2},  \nonumber  \\
\psi^{8,0}_6&=&(\ket{022}-\ket{202})/\sqrt{2},  \nonumber  \\
\psi^{8,0}_7&=&(-\ket{021}-\ket{120}+\ket{201}+\ket{210})/2,  \nonumber  \\
\psi^{8,0}_8&=&(2\ket{012}+\ket{021}-2\ket{102} \nonumber \\
             &&-\ket{120}-\ket{201}+\ket{210})/\sqrt{12}.
\eea
The second octet of states carries a degeneracy label $1$ 
and forms the logical one state $\ket{1_L}$, 
\begin{eqnarray}
\psi^{8,1}_1\!&=&\!\!(2\ket{002}-\ket{020}-\ket{200})/\sqrt{6}  \nonumber \\
\psi^{8,1}_2\!&=&\!\!(2\ket{001}-\ket{010}-\ket{100})/\sqrt{6}  \nonumber \\
\psi^{8,1}_3\!&=&\!\!(-2\ket{110}+\ket{011}+\ket{101})/\sqrt{6} \nonumber  \\
\psi^{8,1}_4\!&=&\!\!(2\ket{112}-\ket{121}-\ket{211})/\sqrt{6}  \nonumber \\
\psi^{8,1}_5\!&=&\!\!(-2\ket{221}+\ket{122}+\ket{212})/\sqrt{6}  \nonumber \\
\psi^{8,1}_6\!&=&\!\!(-2\ket{220}+\ket{022}+\ket{202})/\sqrt{6}  \nonumber \\
\psi^{8,1}_7\!&=&\!\!(2\ket{012}-\ket{021}+2\ket{102} \nonumber  \\
             &&\; -\ket{120}-\ket{201}-\ket{210})/\sqrt{12}  \nonumber \\
\psi^{8,1}_8\!&=&\!\!(-\ket{021}+\ket{120}-\ket{201}+\ket{210})/2. 
\end{eqnarray}
In other words, the first superscript denotes the dimension of 
the representation, the second is a degeneracy label and the 
subscript labels the state within the representation.  

As in the 
case of the three-qubit DFS/NS, the logical zero state is given 
by $\ket{0_L}= \sum_i\alpha_i\psi^{8,0}_i$ (arbitrary superposition) 
and likewise for $\ket{1_L}= \sum_i\beta_i\psi^{8,1}_i$.  
Using the notation of Sec.~\ref{sec:3qbdfs}, the logical states 
can be fully identified by the quantum numbers, 
$\ket{p,q;\lambda;t,t_3,y}$, where $p,q$ are the principal 
quantum numbers which identify the irreducible representation, 
$\lambda$ is the degeneracy label, and 
$t,t_3,y$ identify the states within the representation 
by its secondary quantum numbers.  
The states of the octet which comprise the logical zero 
state are, in this notation, given by
\bea
\psi^{8,0}_1\!&=&\!\! \ket{1,1;0;1,1,0},\nonumber \\
\psi^{8,0}_2\!&=&\!\! \ket{1,1;0;1/2,1/2,1},\nonumber \\
\psi^{8,0}_3\!&=&\!\! \ket{1,1;0;1/2,-1/2,1},\nonumber \\
\psi^{8,0}_4\!&=&\!\! \ket{1,1;0;1,-1,0},\nonumber \\
\psi^{8,0}_5\!&=&\!\! \ket{1,1;0;1/2,-1/2,-1},\nonumber \\
\psi^{8,0}_6\!&=&\!\! \ket{1,1;0;1/2,1/2,-1},\nonumber \\
\psi^{8,0}_7\!&=&\!\! \ket{1,1;0;1,0,0},\nonumber \\
\psi^{8,0}_8\!&=&\!\! \ket{1,1;0;0,0,0}. 
\eea
The states which comprise the logical one are given by
\begin{eqnarray}
\psi^{8,1}_1\!&=&\!\! \ket{1,1;1;1,1,0}, \nonumber \\
\psi^{8,1}_2\!&=&\!\! \ket{1,1;1;1/2,1/2,1}, \nonumber \\
\psi^{8,1}_3\!&=&\!\! \ket{1,1;1;1/2,-1/2,1}, \nonumber \\
\psi^{8,1}_4\!&=&\!\! \ket{1,1;1;1,-1,0}, \nonumber \\
\psi^{8,1}_5\!&=&\!\! \ket{1,1;1;1/2,-1/2,-1}, \nonumber \\
\psi^{8,1}_6\!&=&\!\! \ket{1,1;1;1/2,1/2,-1}, \nonumber \\
\psi^{8,1}_7\!&=&\!\! \ket{1,1;1;1,0,0}, \nonumber \\
\psi^{8,1}_8\!&=&\!\! \ket{1,1;1;0,0,0}.
\end{eqnarray}
The remaining 11 states include a (completely antisymmetric) 
singlet
\beq
\psi_s=(\ket{012}-\ket{021}-\ket{102}+\ket{120}-\ket{210})/\sqrt{6} 
\eeq
and a (completely symmetric) decuplet of states:
\bea
\psi_1^{10} &=& \ket{111}, \nonumber \\
\psi_2^{10} &=& (\ket{011}+\ket{101}+\ket{110})/\sqrt{3},\nonumber \\
\psi_3^{10} &=& (\ket{001}+\ket{010}+\ket{100})/\sqrt{3},\nonumber \\
\psi_4^{10} &=& \ket{000}, \nonumber \\
\psi_5^{10} &=& (\ket{112}+\ket{121}+\ket{211})/\sqrt{3},\nonumber \\
\psi_6^{10} &=& (\ket{012}+\ket{021}+\ket{102}+\ket{120}+\ket{210})/\sqrt{6}, 
\nonumber \\
\psi_7^{10} &=& (\ket{002}+\ket{020}+\ket{200})/\sqrt{3},\nonumber \\
\psi_8^{10} &=& (\ket{122}+\ket{212}+\ket{221})/\sqrt{3},\nonumber \\
\psi_9^{10} &=& (\ket{022}+\ket{202}+\ket{220})/\sqrt{3},\nonumber \\
\psi_{10}^{10} &=& \ket{222}.
\eea

A basis for the collective errors for qutrit states is given 
by sums of the operators $\{\lambda_i\}$ of Appendix\ref{app:alg}:
\beq
S^\alpha= \sum_i\lambda^\alpha_i,
\eeq
where $\alpha=1,2,...,8$ and $i$ denotes the 
physical qutrit $1,2$ or $3$.  A generic collective error 
has the form 
\beq
S = \sum_\alpha a_\alpha S^\alpha,
\eeq
where the $a_\alpha$ are arbitrary constants.  As in the 
three qubit DFS/NS, the states within blocks 
$\ket{0_L}$ and $\ket{1_L}$ get mixed with each other 
in exactly the same way during collective operations, 
but states in one block are not mixed with states in 
another.  Logical operations will mix these blocks 
with each other.  In the DFS/NS basis, the operators 
$S_{\mbox{\scriptsize dfs}}^\alpha %
=V_{\mbox{\scriptsize dfs}}^{\phantom{-1}}S^\alpha %
V_{\mbox{\scriptsize dfs}}^{-1}$ 
are block diagonal in accordance with 
Eq.~(\ref{eq:dfsmatrix}).  Let us order the states in 
a column vector: $\Psi$ = column$\{$$\psi^{8,0}_1$,
$\psi^{8,0}_2$,
$\psi^{8,0}_3$,
$\psi^{8,0}_4$,
$\psi^{8,0}_5$,
$\psi^{8,0}_6$,
$\psi^{8,0}_7$,
$\psi^{8,0}_8$,
$\psi^{8,1}_1$,
$\psi^{8,1}_2$,
$\psi^{8,1}_3$,
$\psi^{8,1}_4,$
$\psi^{8,1}_5,$
$\psi^{8,1}_6,$
$\psi^{8,1}_7,$
$\psi^{8,1}_8,$
$\psi_s,$
$\psi_1^{10},$
$\psi_2^{10},$
$\psi_3^{10},$
$\psi_4^{10},$
$\psi_5^{10},$
$\psi_6^{10},$
$\psi_7^{10},$
$\psi_8^{10},$
$\psi_9^{10},$
$\psi_{10}^{10}$
$\}$.
From these states one may readily deduce the transformation 
$V_{\mbox{\scriptsize dfs}}$ which 
takes the qutrit computational basis states to the 
DFS/NS basis.  It is then clear that $V_{\mbox{\scriptsize dfs}}$ 
is a $27\times 27$ matrix of $SU(3)$ 
Wigner-Clebsch-Gordan coefficients.  
Since, the collective errors in this basis are block diagonal 
[viz. Eq.~(\ref{eq:dfsmatrix})], these blocks will 
be labeled according 
to the set of states on which they act nontrivially.  Let 
$S_0$ be the first such block (which acts nontrivially 
on the states which form the logical zero), 
$S_1$ be the second such block (which acts nontrivially 
on the states which form the logical one), 
$S_s$ be the third such block (which acts 
on the singlet state), and $S_{10}$ which acts 
on the states in the decuplet.  The form of the matrix 
$S_{\mbox{\scriptsize dfs}} =%
V_{\mbox{\scriptsize dfs}}^{\phantom{-1}}SV_{\mbox{\scriptsize dfs}}^{-1}$ 
is given by 
\beq
\label{eq:coll33errs}
S_{\mbox{\scriptsize dfs}} = 
                   \left(\begin{array}{cccc}
		     S_0 &0 &0 &0 \\
		      0 & S_1 &0 &0 \\
		      0 &0 & S_s &0 \\
		      0 &0 &0 & S_{10}
		              \end{array}\right),
\eeq
where $S_0$ and $S_1$ are both $8 \times 8$ matrices and are given by
\begin{widetext}
\beq
\left(\begin{array}{cccccccc} 
      2a_3 & a_6+ia_7 & 0 & 0 & 0 & a_4-ia_5 & \sqrt{2}(a_1-ia_2) & 0 \\
 a_6-ia_7 & a_3+\sqrt{3}a_8 & a_1-ia_2 & 0 & 0 & 0 
& \frac{-a_4+ia_5}{\sqrt{2}}& \frac{-3(a_4-ia_5)}{\sqrt{6}}\\
0& a_1+ia_2& -a_3+\sqrt{3}a_8 & -a_4+ia_5 &0& 0 &\frac{a_6-ia_7}{\sqrt{2}}&
\frac{-3(a_6-ia_7)}{\sqrt{6}} \\
0 & 0 &-a_4-ia_5 & -2a_3 &a_6-ia_7 & 0 & \sqrt{2}(a_1+ia_2) & 0 \\
0 & 0 & 0 &a_6+ia_7 & -a_3-\sqrt{3}a_8 & a_1+ia_2& \frac{a_4+ia_5}{\sqrt{2}}& \frac{3(a_4+ia_5)}{\sqrt{6}} \\
a_4+ia_5 & 0 & 0 & 0 & a_1-ia_2 & a_3-\sqrt{3}a_8 & \frac{a_6+ia_7}{\sqrt{2}} &\frac{-3(a_6+ia_7)}{\sqrt{6}} \\
\sqrt{2}(a_1+ia_2) & \frac{-a_4-ia_5}{\sqrt{2}}&\frac{a_6+ia_7}{\sqrt{2}} 
&\sqrt{2}(a_1-ia_2)&\frac{a_4-ia_5}{\sqrt{2}}&\frac{a_6-ia_7}{\sqrt{2}}&0&0\\
0 &\frac{-3(a_4+ia_5)}{\sqrt{6}} &\frac{-3(a_6+ia_7)}{\sqrt{6}} & 0 & \frac{3(a_4-ia_5)}{\sqrt{6}} & \frac{-3(a_6-ia_7)}{\sqrt{6}} & 0 & 0 
		  \end{array}\right). 
\eeq
\end{widetext} 
The matrix $S_s$ is a $1\times 1$ zero ``matrix'' and the 
$10 \times 10$ matrix $S_{10}$ will not be displayed since it 
is not relevant for the DFS/NS.  

In summary, if the physical circumstances are such that the 
errors/operations on a set of three qutrits in the $\3$ 
representation are identical on each qutrit, then errors/operators 
will have the form, of Eq.(\ref{eq:coll33errs}).  There is then 
a two-state subsystem formed by two collections of states 
$\psi^{8,0}_i$ and $\psi^{8,1}_j$ which may represent a 
decoherence-free, or noiseless, subsystem.


\section{Discussion and Conclusions}

\label{sec:concl}

We note that this research was prompted, in part, 
by the following question:
How does a quantum state or operator transform?  This 
is a fundamental, physically motivated question.  
The transformation properties determine 
the good quantum numbers of a state.  For three-state, 
and higher-dimensional systems, the states could 
transform in one of two ways under special unitary 
transformations and the representations are 
inequivalent.  There are several physical 
consequences of the difference in transformation 
properties.  

For the theory of decoherence-free, or 
noiseless, subsystems it is important to determine 
the transformation properties which distinguish different 
physical states.  Without this knowledge, 
it is not possible to reliably form a DFS/NS.  Using 
Weyl's unitary trick, this is clearly seen through the use of 
Young's tableau for analyzing the irreps of the $SU(d)$ groups.

Similarly, simulating quantum systems with other quantum 
systems requires strict adherence to the appropriate transformation 
rules during the applications of quantum controls.  A very 
important example of this is provided by low-energy nuclear 
interactions and the quark model.  Quark-quark interactions 
at low energies involve both weak and strong forces.  Both 
the theories of strong and weak forces 
involve $SU(d)$ symmetry groups.  QCD is 
a non-Abelian gauge theory, with gauge group $SU(3)$.  In 
this theory, quarks transform according to the $\3$ rep 
of the group $SU(3)$.  Weak interaction physics, 
or ``flavor'' physics, potentially involves six flavors of 
quarks which have an approximate $SU(6)$ symmetry.  At lower 
energies, only the quarks with mass less than the 
experimental interaction energies are used in calculations.  
The three lightest quarks are quite close in mass and 
have an approximate $SU(3)$ symmetry to an even 
better approximation than the $SU(6)$ theory.  (Reference 
\cite{Weinberg:QTFII} contains detailed discussions 
of these topics.)  Whereas quarks transform according 
to the $\3$ rep, 
antiquarks transform according to the $\tb$ rep.  Baryons, 
such as the proton and neutron, are color-neutral bound 
states of three quarks.  Mesons behave as color-neutral 
states of quarks and antiquarks.  
Thus the transformation 
properties of the particles involved in low energy 
nuclear interactions are critically important in simulations.  
Since there exist states of particles which behave in 
such a way,the differences in transformation properties 
of quantum systems must be taken into account during the 
simulation of low-energy nuclear physics.  

We may therefore conclude that the existence of inequivalent 
fundamental irreps for $SU(d)$ can be vital for quantum 
information processing, whether the systems being 
used to process quantum information contain $d$ distinct 
orthogonal states, or a system being simulated contains 
$d$ such states.  
Clearly there is a great deal of work still to be done in this 
area.  Whether or not a system transforms according to a barred 
or unbarred representation is determined by the physical system.  
Not all systems will naturally 
obey a super-selection rule of this sort.  

In the near future, we anticipate exploring quantum 
computing in DFS/NSs constructed from these higher-dimensional 
state spaces.  We also expect to more fully explore the 
experimental circumstances which give rise 
to DFS/NSs and how one would control the system to keep it 
in a DFS/NS.


\begin{acknowledgments}

The author thanks ORDA of Southern Illinois University for 
partial support of this work under internal grant 
4-14095 and Centro de Ci\^{e}ncias Matem\'{a}ticas at the University 
of Madeira.  This work was supported in part by POPRAM III and CITMA, 
Portugal, and was undertaken during the XXIX 
Madeira Math encounter.  The author also thanks V. Akulin, 
J. Clark, A. Mandilara, and especially N. Harshman for stimulating and 
helpful discussions.  

\end{acknowledgments}



\appendix

\section{Group Theory and Young tableau}

\label{app:gpth}


\subsection{Group theory primer}

In this appendix various definitions are collected which may 
or may not be familiar to the reader.  Since 
only matrix representations are considered here, one should place 
the word ``matrix'' in front of ``representation'' throughout the 
article to be precise.  

If there exists a homomorphic mapping of a group ${\cal G}$ onto a group 
of nonsingular $d\times d$ matrices $\Gamma(T) $, for all 
$T\in {\cal G}$ then the set of matrices $\Gamma(T)$ forms a 
$d$-dimensional (matrix) {\it representation} $\Gamma$ of the 
group ${\cal G}$.  
If the homomorphism is also an isomorphism (one to one {\it and} 
onto) then the representation is said to be {\it faithful}.  
Here representations of $SU(d)$ will be considered.  
The lowest-dimensional set of matrices which faithfully represents 
it contains $d\times d$ matrices.  
A faithful representation of 
$SU(d)$ by $d\times d$ matrices is called a {\it fundamental} 
representation and is necessarily irreducible.
(We will take 
``irreducible representation'' to mean that the representation cannot 
be written in terms of matrices of smaller dimension \cite{irrednote}.)
Two representations, $\Gamma$ and $\Gamma^\prime$, of a group ${\cal G}$ 
are said to be {\it equivalent}, if there exists a similarity 
transformation $S$ such that 
$$
\Gamma^\prime = S\Gamma S^{-1},
$$
for all $\Gamma \in {\cal G}$.  If there is no such transformation, 
then the two representations are said to be {\it inequivalent}.

{\bf Theorem} There are two inequivalent irreducible fundamental 
representations of $SU(d)$ for $d\geq 3$.


\subsection{Young's tableau}

Here, a brief summary of some properties of Young tableau 
are given.  

Young's tableau enable the determination of the irreducible 
components of a tensor product.  These methods are used in the text 
to find the direct sum of irreps 
arising from a tensor product 
of irreps.  For a more detailed explanation and derivation 
of the rules, consult a text on group theory-for example 
\cite{Biedenharn,Cornwell:84II}.  
Since only $SU(d)$ representations (and corresponding equivalent 
algebraic representations) are used in this article, the rules 
are discussed in terms of $SU(d)$ irreps.  

For representations of $SU(N)$, a single box
$
\setlength{\arrayrulewidth}{.4pt}
\begin{tabular}{|c|}
\hline \phantom{ai}\\ \hline
\end{tabular}
$
corresponds to a fundamental irreducible representation.  The 
dimension of the representation is determined by filling the 
box with numbers $1,2,...,N$. 

A set of two boxes corresponding to two 
antisymmetric indices 
under their interchange is provided by vertical boxes:
$
\setlength{\arrayrulewidth}{.4pt}
\begin{tabular}{|c|}
\hline \phantom{ai} \\ \hline
 \phantom{ai} \\ \hline 
\end{tabular}.
$
Similarly, two horizontal boxes, 
$
\setlength{\arrayrulewidth}{.4pt}
\begin{tabular}{|c|c|}
\hline \phantom{ai} &
 \phantom{ai} \\ \hline 
\end{tabular},
$
represent indices which are symmetric under interchange.
In general, the indices must be antisymmetric under interchange of 
numbers in columns of the tableau and symmetric in the rows.  

To determine the number of states in an irrep, the 
boxes are filled with numbers.  
For example there are 
$N$ states in the fundamental irrep of $SU(N)$.  Thus 
different integers $1,2,...,N$ can be put into 
the boxes, each set of numbers in the boxes representing 
a different state within the irrep.  
In order to prevent overcounting, the boxes are filled such 
that the numbers are nondecreasing from left to right.  Filling 
boxes which are above one another requires the numbers 
to be distinct for a nonzero tableau due to the antisymmetry.  
In addition, overcounting will be prevented 
if the numbers are filled in a strictly increasing order from top 
to bottom.  

A familiar example is provided in Sec.~\ref{sec:3qbdfs} 
where the product of two two-state systems is shown to produce 
a singlet and a triplet using Young's Tableau.


\subsection{Singlet states}

\label{app:singlets}

Here, a set of raising and lowering operators are given 
for $SU(3)$ in terms of the Gell-Mann matrices.  It is also shown 
that the state $\ket{\phi^\prime} = (1/\sqrt{3})\sum_i\ket{i\bar{i}}$ 
is a singlet state.

There are several ways to show that $\ket{\phi^\prime}$ is a 
singlet state.  To do it directly from relations 
(\ref{eq:singletstaterl}), one 
can use the relations for the raising and lower operators on 
the states.  In terms of the Gell-Mann matrices,
\bea
\label{eq:rnlops}
T_\pm= \half(\lambda_1\pm i\lambda_2), && T_3 = \half\lambda_3, 
                \nonumber  \\
V_\pm= \half(\lambda_4 \pm i\lambda_5), 
                && U_\pm = \half(\lambda_6 \pm i\lambda_7), 
                 \nonumber  \\
Y = \frac{1}{\sqrt{3}}\lambda_8.\;\;\;\;\;\;\;\;\;\;&&
\eea
One may also define $V_3=(1/2)(\sqrt{3}\lambda_8+T_3)$ 
and $U_3=(1/2)(\sqrt{3}\lambda_8-T_3)$.  This proves convenient 
since one may use the known raising and lowering operation 
relations of $SU(2)$; $\lambda_4,\lambda_5,V_3$ and 
$\lambda_6,\lambda_7,U_3$ each form an $SU(2)$ algebra.  
For a given $p,q$, the states of an irrep may be labeled using 
$t,$ $t_3,$ and $y$.  The states are eigenstates of $T_3$ and $Y$ 
by construction,
$$
T_3\ket{t,t_3,y} = t_3\ket{t,t_3,y}, \;\;\;\; 
    Y\ket{t,t_3,y} = y\ket{t,t_3,y}.
$$
At this point one would proceed essentially as is done for 
finding the raising and lowering operator relations for $SU(2)$.  
However, this is somewhat tedious and there is an easier way.  

A singlet representation transforms trivially under actions of 
the group, as a scalar.  Therefore, if one notes that the qutrit 
states transform according to the representation $U$, 
for $i = 0,1,2$,  
$$
\ket{i}\rightarrow 
    \sum_j U_{ij}\ket{j}, \;\; \ket{i}\in {\3},
$$
and the barred states transform according to the representation 
$U^*$, the complex conjugate of $U$, 
$$
\ket{\bar{i}}\rightarrow 
     \sum_{k}U^*_{ik}\ket{\bar{k}}, \;\; \ket{\bar{i}}\in {\tb};
$$
then, under the transformation of each, 
\bea
\sum_i \ket{i}\ket{\bar{i}} &\rightarrow & 
     \sum_{ijk}U_{ij}\ket{j}U^*_{ik}\ket{\bar{k}} \nonumber \\
         && = \sum_{ijk}U_{ij}(U^\dagger)_{ki}\ket{j}\ket{\bar{k}}
                   \nonumber \\
	 && = \sum_{jk}\delta_{jk}\ket{j}\ket{\bar{k}} 
		   \nonumber \\
	 && = \sum_k \ket{k}\ket{\bar{k}}.
\eea
Therefore the state $ \sum_k \ket{k}\ket{\bar{k}}$ 
is an $SU(3)$ scalar.  It is not difficult to convince oneself 
that this implies that the raising and lowering operators acting 
on this state give zero.  (Yet another proof uses the 
explicit parameterization and differential operators 
found in \cite{Byrd/Sudarshan}.)  
The eigenvalues of $T_3$ and $Y$ are 
clearly zero since these act as derivations, 
$$
{\cal O} \sum_i \ket{i\bar{i}} = 
   \sum_i({\cal O} \ket{i})  \ket{\bar{i}}+ \ket{i}({\cal O} \ket{\bar{i}}) 
                       = 0,
$$
and for any operator in the Lie algebra and the eigenvalues 
of barred states are opposite those of the unbarred states.


\section{The algebra of $SU(3)$}

\label{app:alg}

In this appendix the Gell-Mann matrices are listed.  This is one 
basis for the algebra which is commonly used for the 
subgroup chain that is 
discussed throughout the article.  However, it should be emphasized 
that this is not the only basis one could choose.  
The Casimir operators are also given for $SU(3)$ since they are 
elements of the CSCO discussed in the text.  

The Gell-Mann matrices are given by 
$$
\lambda_1 = \left( \begin{array}{crcl}
                     0 & 1 & 0 \\
                     1 & 0 & 0 \\
                     0 & 0 & 0   \end{array} \right), \;
\lambda_2 = \left( \begin{array}{crcr}
                     0 & -i & 0 \\
                     i &  0 & 0 \\
                     0 &  0 & 0   \end{array} \right), 
$$
$$
\lambda_3 =  \left( \begin{array}{crcr}
                     1 &  0 & 0 \\
                     0 & -1 & 0 \\
                     0 &  0 & 0   \end{array} \right), \;
\lambda_4 =  \left( \begin{array}{clcr}
                     0 & 0 & 1 \\
                     0 & 0 & 0 \\
                     1 & 0 & 0   \end{array} \right), 
$$
$$
\lambda_5 = \left( \begin{array}{crcr}
                     0 & 0 & -i \\
                     0 & 0 & 0 \\
                     i & 0 & 0   \end{array} \right), \;
 \lambda_6 = \left( \begin{array}{crcr}
                     0 & 0 & 0 \\
                     0 & 0 & 1 \\
                     0 & 1 & 0   \end{array} \right), 
$$
$$
\lambda_7 = \left( \begin{array}{crcr}
                     0 & 0 & 0 \\
                     0 & 0 & -i \\
                     0 & i & 0   \end{array} \right), \;
\lambda_8 = \frac{1}{\sqrt{3}}\left( \begin{array}{crcr}
                     1 & 0 & 0 \\
                     0 & 1 & 0 \\
                     0 & 0 & -2   \end{array} \right).
$$

The Casimir operators for $SU(3)$ are proportional to the operator 
which is quadratic in the algebraic elements,
\begin{equation}
C_2 = \sum_{i=1}^8 \lambda_i^2,
\end{equation}
and the operator which is cubic in the algebraic elements,
\begin{equation}
C_3 = \sum_{i=1}^8 d_{ijk} \lambda_i\lambda_j\lambda_k,
\end{equation}
where 
$$
d_{ijk} = \frac{1}{4}\tr([\lambda_i\lambda_j+\lambda_j\lambda_i]\lambda_k).
$$
These can be written in terms of functions of $p$ and $q$ using 
the operators $T_3$, $Y$ and the raising and lower operators 
\cite{symmetry:book}.



\begin{thebibliography}{53}
\bibitem[{oct()}]{octetnote}
\bibinfo{note}{{This is a famous product in particle physics which prompted the
  name ``Eightfold way" \cite{Gell-Mann:64}}}.
\bibitem[{irr()}]{irrednote}
\bibinfo{note}{The formal definition has been omitted here. See for example,
  \cite{Cornwell:84I+II}, pages 76-83.}
\end{thebibliography}


\end{document}